

\input phyzzx

\def\ea{{\it et al.\/}}
\def\noblackbox{\overfullrule=0pt}
\noblackbox

\Pubnum={UFIFT-HEP-92-8}
\titlepage
\title{MASS AND MIXING ANGLE PATTERNS \break
IN THE STANDARD MODEL \break
AND ITS MINIMAL SUPERSYMMETRIC EXTENSION%
\footnote\dag{This work has been supported in part by the United States %
Department of Energy under contract DE-FG05-86ER-40272.}}
\bigskip
\author{H.~Arason, D.~J.~Casta\~no,
E.~J.~Piard, and P.~Ramond}

\address{Institute for Fundamental Theory\break
Department of Physics, University of Florida\break
Gainesville, Florida 32611 USA}

\bigskip
\abstract{
Using renormalization group techniques, we examine several interesting
relations among masses and mixing angles of quarks and leptons in the Standard
Model.  We extend the analysis to the minimal supersymmetric extension to
determine its effect on these mass relations.  Remarkably Supersymmetry allows
for these relations to be satisfied at a single grand unified scale.}

\submit{Physical Review D}
\endpage

\def\PRL#1&#2&#3&{\ Phys. Rev. Lett.\ \bf #1\rm ,\ #2\ (19#3)}
\def\PRD#1&#2&#3&{\ Phys. Rev. D\ \bf #1\rm ,\ #2\ (19#3)}
\def\PR#1&#2&#3&{\ Phys. Rev.\ \bf #1\rm ,\ #2\ (19#3)}
\def\ZPC#1&#2&#3&{\ Z. Phys. C\ \bf #1\rm ,\ #2\ (19#3)}
\def\NPB#1&#2&#3&{\ Nucl. Phys.\ \bf B#1\rm ,\ #2\ (19#3)}
\def\PLB#1&#2&#3&{\ Phys. Lett.\ \bf #1B\rm ,\ #2\ (19#3)}
\def\CNPP#1&#2&#3&{\ Comments Nucl. Part. Phys.\ \bf #1\rm ,\ #2\ (19#3)}
\def\NC#1&#2&#3&{\ Nuovo Cim.\ \bf #1\rm ,\ #2\ (19#3)}
\def\RMP#1&#2&#3&{\ Rev. Mod. Phys.\ \bf #1\rm ,\ #2\ (19#3)}
\def\MPLA#1&#2&#3&{\ Mod. Phys. Lett.\ \bf A#1\rm ,\ #2\ (19#3)}
\def\PTP#1&#2&#3&{\ Prog. Theor. Phys.\ \bf #1\rm ,\ #2\ (19#3)}
\def\AnnPhysNY#1&#2&#3&{\ Ann. Phys.(NY)\ \bf #1\rm ,\ #2\ (19#3)}
\def\AnnMath#1&#2&#3&{\ Ann. Math.\ \bf #1\rm ,\ #2\ (19#3)}
\def\TransMath#1&#2&#3&{\ Trans. Amer. Math. Soc. \ \bf #1\rm ,\ #2\ (19#3)
}
\def\SJNP#1&#2&#3&{\ Sov. J. Nucl. Phys.\ \bf #1\rm ,\ #2\ (19#3)}
\def\PREP#1&#2&#3&{\ Phys. Reports\ \bf #1\rm ,\ #2\ (19#3)}
\def\NCL#1&#2&#3&{\ Nuovo Cim. Lett.\ \bf #1\rm ,\#2\ (19#3)}
\def\CMP#1&#2&#3&{\ Commun. Math. Phys.\ \bf #1\rm ,\ #2\ (19#3)}
\def\JMP#1&#2&#3&{\ J. Math. Phys.\ \bf #1\rm ,\ #2\ (19#3)}
\def\TMP#1&#2&#3&{\ Theor. Math. Phys.\ \bf #1\rm ,\ #2\ (19#3)}
\def\ApJ#1&#2&#3&{\ Astrophys. J.\ \bf #1\rm ,\ #2\ (19#3)}

\REF{\GL}{M.~K.~Gaillard and B.~Lee, \PRD 10&897&74&.}
\REF{\GUT}{J.~C.~Pati and A.~Salam, \PRD 10&275&74&.}
\REF{\GG}{H.~Georgi and S.~L.~Glashow, \PRL 32&438&74&.}
\REF{\GQW}{H.~Georgi, H.~Quinn, and S.~Weinberg, \PRL 33&451&74&.}
\REF{\amaldi}{U.~Amaldi, W.~de~Boer, and H.~F\"urstenau, \PLB 260&447&91&;
     J.~Ellis, S.~Kelley, and D.~Nanopoulos, \PLB 260&131&91&;
     P.~Langacker and M.~Luo,\PRD 44&817&91&.}
\REF{\nilles}{For a review see H.~P.~Nilles, \PREP 110C&1&84& and references
 therein.}
\REF{\gmrs}{M.~Gell-Mann, P.~Ramond, and R.~Slansky, in {\it Supergravity}
            edited by P.~Van Nieuwenhuizen and D.~Freedman
            (North-Holland, Amsterdam, 1979)}
\REF{\zee}{See reprints in A.~Zee, {\it The Unity of Forces in the Universe},
 (World Scientific, Singapore, 1982), Vol. I, p. 310-335.}
\REF{\mesh}{H.~Harari, H.~Haut, and J.~Weyers, \PLB 78&459&78&;
 Y.~Koide, \PRD 28&252&83&;
 P.~Kaus and S.~Meshkov, \MPLA 3&1251&88&.}
\REF{\ros}{J.~L.~Rosner and M.~Worah, EFI Report No. EFI 92-12 (unpublished).}
\REF{\buras}{M.~S.~Chanowitz, J.~Ellis, and M.~K.~Gaillard,
 \NPB 128&506&77&; A.~Buras, J.~Ellis, M.~K.~Gaillard, and D.~Nanopoulos,
 \NPB 135&66&78&.}
\REF{\GJ}{H.~Georgi and C.~Jarlskog, \PLB 89&297&79&.}
\REF{\okgto}{R.~Gatto, G.~Sartori, and M.~Tonin, \PLB 28&128&68&;
 R.~J.~Oakes, \PLB 29&683&69&; \PLB 31&620&70& (erratum); \PLB 30&262&70&. }
\REF{\HF}{S.~Weinberg, in {\it A Festschrift for I.~I.~Rabi} [Trans. N. Y.
 Acad. Sci., Ser. II (1977), v. 38], p. 185;
 F.~Wilczek and A.~Zee, \PLB 70&418&77&;
 H.~Fritzsch, \PLB 70&436&77&; \PLB 73&317&78&; \PLB 166&423&86&.}
\REF{\soten}{H.~Fritzsch and P.~Minkowski, \AnnPhysNY 93&193&75&;
             H.~Georgi, in {\it Particles and Fields-1974}, edited by
  C.~E.~Carlson, AIP Conference Proceedings No. 23 (American Institute of
  Physics, New York, 1975) p. 575}
\REF{\HRR}{J.~Harvey, P.~Ramond, and D.~Reiss, \PLB 92&309&80&,
           \NPB 199&223&82&.}
\REF{\LS}{G.~Lazarides and Q.~Shafi, \NPB 350&179&91&.}
\REF{\DHR}{S.~Dimopoulos, L.~Hall, and S.~Raby, OSU Reports No.
           DOE-ER-01545-566, 567 (unpublished); V.~Barger, M.~S.~Berger,
 T.~Han, and M.~Zralek, Wisconsin Report No. MAD/PH/693.}
\REF{\haukur}{H.~Arason, D.~J.~Casta\~no, B.~Keszthelyi, S.~Mikaelian,
 E.~J.~Piard, P.~Ramond, and B.~D.~Wright, in preparation.}
\REF{\ufgs}{H.~Arason, D.~J.~Casta\~no, B.~Keszthelyi, S.~Mikaelian,
 E.~J.~Piard, P.~Ramond, and B.~D.~Wright, Florida Report No. UFIFT-HEP-91-33
 (unpublished).}
\REF{\ARASON}{H.~Arason, D.~J.~Casta\~no, B.~Keszthelyi, S.~Mikaelian,
  E.~J.~Piard, P.~Ramond, and B.~D.~Wright,\PRL 67&2933&91&.
  See also Ref. 21.}
\REF{\Anan}{B.~Ananthanarayan, G.~Lazarides, and Q.~Shafi, \PRD 44&1613&91&;
  S.~Kelley, J.~L.~Lopez, and D.~V.~Nanopoulos, \PLB 274&387&92&.}
\REF{\cdf}{F.~Abe \ea, Central Detector Facility (CDF) Collaboration, \PRL
           68&447&92&.}
\REF{\klauder}{For an earlier version see P.~Ramond, Florida Report No.
   UFIFT-HEP-92-04 (unpublished).}
\REF{\gl}{J.~Gasser and H.~Leutwyler, \PREP 87&77&82&.}
\REF{\pendross}{B.~Pendleton and G.~G.~Ross, \PLB 98&291&81&;
                C.~T.~Hill, \PRD 24&691&81&.}
\REF{\ATWOOD}{Atwood, (private communication).}
\REF{\ng}{D.~Ng and Y.~J.~Ng, \MPLA 6&2243&91&.}

\FIG\one{}
\FIG\four{}
\FIG\five{}
\FIG\seven{}
\FIG\eight{}
\FIG\ten{}
\FIG\twelve{}
\FIG\thirteen{}
\FIG\fourteen{}
\FIG\fifteen{}

\chapter{Introduction}

Most of the parameters in the Standard Model are to be found in the Yukawa
sector of the theory where they parametrize quark and lepton masses,
the interfamily mixings of the quarks, and CP violation. Historically, only
one of these thirteen parameters was ever predicted,\refmark{\GL} the charmed
quark mass, but only after an inspired guess on the value of a strong ({\it
i.e.}, presently incalculable) matrix element.

Theoretical guesses on the nature of physics beyond the Standard Model
abound in the literature. Many use as inspiration the idea of a Grand Unified
Theory\refmark{\GUT,\GG} (GUT) which emerged from the observed pattern of the
quantum numbers of the elementary particles. When applied in conjunction with
the renormalization group\refmark{\GQW}, this idea has proven extremely
fruitful. Recent work indicates that the experimental values of the gauge
couplings are such that all three couplings evolve to the same
value\refmark{\amaldi} at shorter distances only when Supersymmetry is included
at $1-10$ TeV. Without Supersymmetry, the gauge couplings meet two at a time,
forming a small ``GUT triangle'' in the plot of their evolution as a function
of scale.

This encouraging situation, hinting at a Supersymmetric GUT, should
be matched by concomitant simplicity in the other parameters of the theory. To
that purpose we present a comparative analysis of possible relations among
Yukawa couplings at shorter distances both in the Standard Model itself and in
its minimal supersymmetric extension\refmark{\nilles}.

While we find little evidence to support the view that the Standard Model
is by itself the low energy manifestation of a pure GUT, we are encouraged
by the results of this investigation:  Inclusion of Supersymmetry allows many
possible GUT relations among Yukawa couplings to be satisfied at one
appropriate gauge unification scale.
\endpage

\chapter{Models of the Yukawa Sector}

Many models have been proposed to explain the peculiar structure of the Yukawa
couplings. In a certain basis, these Yukawa matrices are well approximated by
the matrix\refmark{\gmrs}
$$
   \pmatrix{0&0&0\cr
            0&0&0\cr
            0&0&1\cr}\ ,
$$
which expresses the fact that one family is so much heavier than the other two.
The theoretical temptation has been to express the masses of the lighter
families and their mixings as radiative effects.\refmark{\zee} Yet this
approach has not yielded any satisfactory models. Another has been to use as
the starting point the ``democratic'' matrix where all entries are equal,
thereby producing two zero eigenvalues. This type of idea could be implemented
by condensates formed by new flavor-blind strong forces.\refmark{\mesh} Yet
another approach is to think of the ``elementary'' particles as composites and
use the Yukawa matrices as hints to infer their structure.\refmark{\ros} In
the following, however, we concentrate on Yukawa matrices that arise naturally
in the context of Grand Unified Theories.

In the context of the $\rm SU(5)$ GUT\refmark{\GG}, several mass relations
arise, based on simple assumptions for the possible Higgs structure. The mass
term for the down quarks and leptons comes from the Yukawa interaction of the
${\overline{\bf 5}}$ and {\bf 10} of fermions.  This leads to relations between
the charge $-1/3$ quarks' and the charged leptons' Yukawa couplings.  With only
a ${\overline{\bf 5}}$ of Higgs, one obtains equality between the $\tau$
lepton and bottom quark masses at the GUT scale:
$$
   m_b=m_\tau\ .
   \eqn\mbmtau
$$
To the level of approximation used at the time, this relation was found to be
consistent at experimental scales, after taking into account the running of the
quark masses.\refmark{\buras}  Similar relations apply to the lighter two
families, but are clearly incompatible with experiment.  To alleviate this, a
new scheme was proposed\refmark{\GJ} with a slightly more complicated Higgs
structure (using a {\bf 45} representation in conjunction with the
${\overline{\bf 5}}$). It replaces the above with the more complicated
relations for the two lighter families
$$
   \eqalign{
   m_d  &= 3m_e\ , \cr
   3m_s &= m_\mu \ . \cr}
   \eqn\geoj
$$

The situation concerning the mixing angles is equally intriguing. There
happens to be a near numerical equality between the
square of the tangent of the Cabibbo angle and the ratio of the down to the
strange quark masses (determined from current algebra). This relation
\refmark{\okgto} reads
$$
   \tan\theta_c\approx {\sqrt{m_d\over m_s}}\ .
   \eqn\tantheta
$$
It has provided the central inspiration in the search for Yukawa matrices.
Very general classes of matrices with judiciously chosen
textures\refmark{\HF} ({\it i.e.\/}, zeroes in the right places) could
reproduce this relation, at least approximately.

In the context of $SO(10)$,\refmark{\soten} these three different relations
could all be obtained in one model,\refmark{\HRR} with the required texture
enforced naturally by discrete symmetries at the GUT scale. In this model the
mixing of the third family with the two lighter ones is dictated exclusively by
the the charge 2/3 quarks' Yukawa matrix. There ensues a relation
for the mixing of the second and third families\refmark{\HRR}
$$
   V_{cb}={\sqrt{m_c\over m_t}}\ ,
   \eqn\vcb
$$
which provides a relation between the top quark mass and the lifetime of the
B meson.

These four relations can all be obtained if one takes the Yukawa mixing
matrices to be of the form\refmark{\HRR} (shown here in a specific basis)
$$
   {\bf Y}_u=\pmatrix{0&P&0\cr P&0&Q\cr0&Q&V\cr}\ ,
$$
\vskip .2cm
$$
   {\bf Y}_d=\pmatrix{0&R&0\cr R&S&0\cr 0&0&T\cr}\ ,~~
   {\bf Y}_e=\pmatrix{0&R&0\cr R&-3S&0\cr 0&0&T\cr}\ .
   \eqn\yukmats
$$
This form has been recently rediscovered by several groups,\refmark{\LS,\DHR}
and some of our analysis overlaps with their work.

Although derived with specific and sometimes complicated Higgs structures in
mind (as in the $SO(10)$ model), these relations may well prove sturdier than
the theories which generated them.  In the following, we first examine the
relations in the context of the Standard Model at varying scales all the way to
Planck scale.  We then extend the analysis to the minimal supersymmetric
extension of the Standard Model, and compare the effect of this extension on
their compatibility at some unified scale.  A more thorough treatment of the
SUSY extension is in preparation.\refmark{\haukur}
\endpage

\chapter{The Renormalization Group}

We outline and will use the numerical techniques and routines developed
in our previous work.\refmark{\ufgs,\ARASON}  We first use experiment to fix
the parameters of the Standard Model at lower energies. We then use these
values as initial conditions in the renormalization group running to shorter
length scales, using the $\overline {\rm MS}$ scheme. In the Standard Model, we
use 2-loop renormalization group equations in evolving the couplings.  In the
supersymmetric extension, we work to 1-loop.  In each case, we include a proper
treatment of thresholds and make no approximations in the Yukawa sector. Our
incomplete knowledge of the Standard Model parameters forces us to repeat the
analysis for a range of allowed values of the top quark and Higgs masses.
Although the top quark mass is not exactly known, it is believed to lie
somewhere between $91$ GeV\refmark{\cdf} and $200$ GeV, the lower limit being
set by direct experimental searches, the upper one by the radiative effect of
the top quark mass on the ratio of neutral to charged current processes
($\rho$-parameter). In these runs, we take $g_3(M_Z)=1.191$ and the physical
bottom quark mass $M_b=4.89$ GeV.

Let us summarize the salient features of the renormalization group running in
the Standard Model. At the one loop level, the gauge couplings are unaffected
by the other couplings in the theory. On the other hand, the Yukawa couplings
are affected at one loop by both the gauge and Yukawa couplings. Since the top
Yukawa coupling is at least as big as the gauge couplings at low energy, that
means the running of the Yukawas is sensitive to mostly the top Yukawa and the
QCD gauge couplings. Thus we can expect the mass and mixing relations we have
just described to be sensitive to the value of the top quark mass. The Higgs
quartic self-coupling enters in the running of the other couplings only at the
two loop level, so that its effect on the quark and lepton parameters is small.
However, its own running is very sensitive to the top quark mass; it can become
negative as easily as it can blow up, corresponding to vacuum instability or to
strong self interaction of the Higgs (triviality bound), respectively. The
discovery of the Higgs with mass outside these bounds would be a signal for
physics beyond the Standard Model. The graphs in Fig.~\one\
summarize these bounds for representative values of the top quark mass.
\bigskip
\centerline{\hbox{
                  \psfig{figure=t100_vac2.ps,height=5.21cm}
                  \psfig{figure=t125_vac2.ps,height=5.21cm}
                 }}
\smallskip
\centerline{\hbox{
                  \psfig{figure=t150_vac2.ps,height=5.21cm}
                  \psfig{figure=t200_vac2.ps,height=5.21cm}
                 }}
\smallskip
\centerline{\singlespace
     \vtop{\parindent=0pt\hsize=5.5truein {\ninerm FIG.~1
Vacuum stability and triviality bounds on the Higgs mass for various {\tenrm
$\scriptstyle M_t$} giving scales of expected new physics beyond the Standard
Model.}}}
\bigskip
\noindent
For example, if $M_t=150$ GeV, we see from the corresponding
plot that a Higgs mass between 95 and 150 GeV need not imply
any new physics up to Planck scale. However, if the Higgs
were observed outside of this range, then some new physics must
appear at the scale indicated by the curve, either because of vacuum
instability if $M_H<95$ GeV or because the Higgs interaction becomes too
strong if $M_H>150$ GeV. It is amusing to note that it is for comparable values
of the top and Higgs masses that these bounds are least restrictive, but it is
important to emphasize that a high value of the top with a relatively low value
of the Higgs necessarily indicates the presence of new physics within reach of
the SSC. Subsequently, when examining the mass and mixing angle relations in
the context of the Standard Model, we will make the choices for $M_t$ and $M_H$
in our renormalization group runs consistent with these bounds.  For a chosen
value of $M_t$, varying $M_H$ within the vacuum stability and triviality bounds
does not affect any of our results, and we will therefore choose a
corresponding, representative value of $M_H$.

As is well known, the Standard Model shows no apparent inconsistencies until
perhaps the Planck scale, where quantum gravity enters the picture. The nature
of the physics to be found between our scale and the Planck scale is a matter
of theoretical taste. At one extreme, because of the values of the gauge
couplings, new phenomena may be inferred every two orders of magnitude. At the
other, there is the possible desert suggested by GUTs; however, the absence of
new phenomena over many orders of magnitude cannot be understood
(perturbatively) unless one generalizes the Standard Model in some way to solve
the hierarchy problem.  Supersymmetrizing the Standard Model at an
experimentally accessible scale can accomplish this. This particular scenario
is bolstered by the fact that with such ``low energy'' Supersymmetry, the three
gauge couplings of the Standard Model meet at one scale ($\sim 10^{16}$ GeV) at
the perturbative value of $\sim 1/26$.\refmark{\amaldi} The collapse of the GUT
triangle in the supersymmetric extension fixes two scales, the one at which the
gauge couplings unify, the other at the threshold of Supersymmetry. Minimal
Supersymmetry implies two Higgs doublets and eliminates the feisty quartic
self-coupling of the Standard Model. But there appears an extra parameter, the
ratio of the vacuum values of these two doublets, parametrized by an angle
$\beta$, $\tan\beta=v_u/v_d$, where $v_u$ ($v_d$) is the vacuum expectation
value of the Higgs field that gives mass to the charge $2/3$ ($-1/3$, $-1$)
fermions. In the following,\refmark{\klauder} we examine certain relations
among masses and mixing angles in the context of the Standard Model itself and
in its minimal supersymmetric extension.  In the latter, we only treat the case
of one light Higgs.

\endpage

\chapter{Running the Relations in the Standard Model}

We now proceed to run these relations using only Standard Model physics.
The three gauge couplings semi-converge in forming the GUT triangle
around $10^{16}$ GeV.

\section{Relation (I): $m_b = m_\tau$}

This relation is the most natural one in the $SU(5)$ theory, and it could be
expected to be valid at scales where the Standard Model gauge couplings are the
closest to one another. We examine its validity for three different physical
values of the top and Higgs masses in the Standard Model. The results are
summarized in Fig.~\four.
\bigskip
\centerline{\hbox{
                  \psfig{figure=mbmtau.ps,height=6.5cm}
                 }}
\smallskip
\centerline{\singlespace
      \vtop{\parindent=0pt\hsize=5.5truein {\ninerm FIG.~2
Plot of $\scriptstyle m_b/m_\tau$ as a function of scale in the Standard Model
for various top and Higgs masses.}}}
\bigskip
\noindent
The noteworthy feature of the figure is that this simplest of the SU(5)
relations is valid at an energy scale many orders of magnitude removed
{}from that at which the gauge couplings tend to converge.\refmark{\ARASON} Our
result is vastly different from that of the original investigations in
Ref.~[\buras]. We have improved on their work by including two loop
effects in the running of the quark Yukawas, by taking into account the full
Yukawa sector, and most importantly by incorporating QCD corrections in the
extraction of the bottom quark mass.\refmark{\ufgs}

\section {Relations (II): $m_d = 3m_e$, $3m_s = m_\mu$}

We now turn to the more complicated relations among masses of the two lighter
families. There are large theoretical uncertainties in the extraction of the
masses of the three lightest quarks from experiment, although the mass ratios
are known more accurately. Following Refs.~[\gl,\ufgs], we take their values to
be $m_d/m_u=1.8$ and $m_s/m_d=21$, so that specifying
$m_s$ fixes $m_d$ and $m_u$.  We note that $m_s/m_d$ and $m_\mu/m_e$
effectively do not run.  Therefore, given this value for $m_s/m_d$, we do
not expect relations (II) to be both satisfied exactly, since
$m_\mu/9m_e \approx 23$.  The uncertainties in the light
quark masses are accounted for by examining the ratios $m_d/3m_e$ and
$3m_s/m_\mu$ for a range of
$m_s(1\ {\rm GeV})$ values from $140$ to $250$ MeV. We have run these same
ratios for representative values of the top and Higgs masses but find the
results to be fairly insensitive to the value of the top.
Therefore, in
Fig.~\five\ , we only present results for top and Higgs masses of $190$ GeV
and $180$ GeV, respectively.
\bigskip
\centerline{\hbox{
                  \psfig{figure=md3me_sm.ps,height=6.5cm} }}
\smallskip
\centerline{\hbox{
                  \psfig{figure=3msmmu_sm.ps,height=6.5cm}
                 }}
\smallskip
\centerline{\singlespace
      \vtop{\parindent=0pt\hsize=5.5truein {\ninerm FIG.~3
Plots of $\scriptstyle m_d/3m_e$ and $\scriptstyle
3m_s/m_\mu$ as a function of scale in the Standard Model for
$\scriptstyle M_t=190$ GeV and $\scriptstyle M_H=180$ GeV.}}}
\bigskip
\noindent
Unlike relation (I) which holds only at $\sim 10^7$ GeV, we see that
relations (II) can hold within $\sim 5\%$ at $10^{16}$ GeV for acceptable
values of the light quark masses.

\section{Relation (III): $\tan\theta_c = \sqrt{m_d/m_s}$}

We find this relation to be quite independent of scale. The reason is that
the Cabibbo angle effectively does not run,\refmark{\ARASON} and the ratio of
light quarks is essentially unaffected by QCD, since both are far away from the
Pendleton-Ross\refmark{\pendross} infrared fixed point. Further, we observe
that their numerical values are fairly independent of the value of the top
quark mass and of the Higgs mass. The agreement is spectacular, hovering
around the 4\% level. For example, for $M_t=100$ GeV and $M_H= 100 $ GeV, we
find that $\tan\theta_c/\sqrt{m_d/m_s}=1.038$ from $M_Z$ to Planck
scale.

\section{Relation (IV): $V_{cb}={\sqrt{m_c/m_t}}$}

This expression involves the top quark mass directly, which may thus be
predicted from this relation.  On the other side of the equation, the
experimental value of the ``$23$'' element of the CKM matrix, $V_{cb}$, is
known only to within $\sim 10\%$,\refmark{\ATWOOD} $V_{cb}=0.043\pm 0.006$.
The value of $V_{cb}$ at all scales is obtained by running the CKM angles.
These numerical results do not depend on the value of the CP-violating phase.

We note that because of the Pendleton-Ross fixed point, the ratio of the two
quark masses runs appreciably in the infrared region. We find that for a top
quark in its lower allowed range, $91-150$ GeV, this relation fails over all
scales.  Accordingly, we present our results for values of $V_{cb}$ and $M_t$
for which the relation can be satisfied below Planck scale. We use for
$V_{cb}$ values ranging from its central value of .043 to .050.  Here and
in the following, we take the value of $m_c$ at $1$ GeV to be $1.41$ GeV.

The results of our runs can be summarized in Fig.~\seven\ in which
we plot both $V_{cb}$ and $\sqrt{m_c/m_t}$ as a function of scale.
\bigskip
\centerline{\hbox{
                  \psfig{figure=vcb178.ps,height=5.21cm}
                 }}
\smallskip
\centerline{\hbox{
                  \psfig{figure=vcb180.ps,height=5.21cm}
                  \psfig{figure=vcb197.ps,height=5.21cm}
                 }}
\centerline{\singlespace
      \vtop{\parindent=0pt\hsize=5.5truein {\ninerm FIG.~4
Plot of $\scriptstyle V_{cb}$ and $\scriptstyle \sqrt{m_c/m_t}$
as a function of scale in the Standard Model for $\scriptstyle M_t=178,\ 180$,
and $\scriptstyle 197$ GeV, for both the central value (.043) and the maximal
value (.050) of $\scriptstyle V_{cb}(M_Z)$.}}}
\bigskip
\noindent
{}From the first plot, we see
that the top quark has to be at least 178 GeV for $\sqrt{m_c/m_t}$ to meet
$V_{cb}$ at the Planck scale. The second plot shows that  $M_t=180$ GeV allows
for this relation to be easily satisfied at $10^{16}$ GeV, if
$V_{cb}(M_Z)=.05$.  This means that a few GeV difference in $M_t$ changes the
meeting of the curves (both of which are affected by $M_t$) by three orders
of magnitude!
Finally, in the third plot, we see that for this relation to be valid
at the unification scale, using the central value of $V_{cb}$, a
197 GeV top quark is needed. We conclude that, given the uncertainties in
the value of $V_{cb}$, this relation may well be valid as long as $M_t>175$
GeV.

\section{Other Possible Mass Relations}

Another interesting mass relation involves
the ratio of the determinants of the charge -1/3 to charge -1 mass matrices
and should equal one if relations (I) and (II) are valid.  We note that,
independent of the top mass ($91$-$200$ GeV), this weaker
(less predictive) relation ($m_d m_s m_b = m_e m_\mu m_\tau$) can be satisfied
at $10^{16}$ GeV for quark masses within the range stated above.

The relations considered so far have been motivated by specific theoretical
models.  There are other relations which are not similarly motivated but
which may nevertheless hint at an underlying unified structure.  In our
search for simple relations among the quark masses, we considered an appealing
``geometric mean'' relation in the up sector, namely
$$
   m_u m_t = m_c^2 \ .
   \eqn\ugeomean
$$
This relation favors higher top quark masses and can be satisfied well at
$10^{16}$ GeV in the $M_t=190$ GeV, $M_H=180$ GeV scenario with an up
quark mass compatible with that needed to satisfy relations (II) at $10^{16}$
GeV.

A similar relation involving the down-type quarks was tested
$$
   m_d m_b = m_s^2 \ .
   \eqn\dgeomean
$$
This relation favors higher top quark masses as well.  In fact, in order to
satisfy Eqs.~\ugeomean\ and~\dgeomean\ at $10^{16}$ GeV, as well as
relations (II), given a fixed value for $m_s(1~{\rm GeV})$ within the range
cited, the top quark mass would have to be larger than $200$ GeV. Fortuitously,
such a value would also favor relation (IV). These geometric mean relations
have been discussed in the literature recently.\refmark{\ng}

To conclude our analysis of the Standard Model case, we see that it is hard to
arrive at a unified picture.
The scale at which relation (I) tends to
be satisfied does not coincide with that at which the others are valid.  Still,
the disagreement is never too large, which makes us hope that small course
corrections in the running of the parameters allow most if not all of these
relations to hold simultaneously at a unified scale.  It is remarkable
that for a top quark at the upper reaches of its allowed range, the long life
of the bottom quark lends plausibility to the $SO(10)$-inspired relation
(IV).
\endpage

\chapter{Running the Relations in the Supersymmetric Standard Model}

\section{Relations (I)-(IV)}

In a previous publication,\refmark{\ARASON} it was shown that low energy
Supersymmetry allows relation (I) to be valid at the scale of gauge
unification.  In fact, if this relation is imposed at the unification scale,
the top quark mass is fixed (up to experimental errors in $g_3$),
for a given value of the angle $\beta$.  The reason is that
the bottom Yukawa running depends both on $g_3$ and on the top Yukawa.
We note that unlike the Standard Model case
the top Yukawa starts increasing at shorter length scales, providing us with
an upper limit on the mass of the top quark. In addition, by relating the
scalar quartic self-coupling below the Supersymmetry breaking scale, to
$\beta$, $g_1$, and $g_2$ above it, the mass of the lightest Higgs is fixed
in terms of the mass of the top (or equivalently $\beta$).
These results are displayed for two scales of Supersymmetry breaking in
Fig.~\eight.
\bigskip
\centerline{\hbox{
                  \psfig{figure=1tev.ps,height=5.21cm}
                  \psfig{figure=10tev.ps,height=5.21cm}
                 }}
\smallskip
\centerline{\singlespace
      \vtop{\parindent=0pt\hsize=5.5truein {\ninerm FIG.~5
Plot of $\scriptstyle M_t$ and $\scriptstyle M_H$ as a function of
$\scriptstyle \beta$. For $\scriptstyle M_{SUSY}=1$ TeV, the high curves
(low curves) correspond to the highest (lowest) value of
$\scriptstyle \alpha_3(M_Z)$ consistent with unification.
For $\scriptstyle M_{SUSY}=10$ TeV, the low and high curves coincide. }}}
\bigskip
\noindent
In the following, we shall take relation (I) as valid at $M_{GUT}$ and assume
the correspondence between $\beta$, $M_t$, and $M_H$ depicted in these figures.
We note, therefore, that this limits the top mass we can consider to
be $\lsim 200$ GeV.
For $M_{SUSY}=1$ TeV, the gauge unification occurs between a low of $6.92\times
10^{15}$ GeV (Low $M_{GUT}$)  and a high of $1.26\times 10^{16}$ GeV
(High $M_{GUT}$), corresponding to $g_3(M_Z)=1.171$ and $g_3(M_Z)=1.197$,
respectively. The error bars in the strong
coupling allow for a SUSY scale as high as $\sim 10$ TeV, with unification at
$6.46\times 10^{15}$ GeV.

The strategy of the remaining part of this paper is to exploit the
relations (II-IV) to constrain $M_t$ and therefore $\beta$ and $M_H$.
For $M_{SUSY}=1$ TeV, we treat two cases,
the first where unification takes place at its lowest value
(Low $M_{GUT}$) and the second where it is at its highest value (High
$M_{GUT}$).  In the following, we will not discuss our results for
the Supersymmetry breaking scale of $10$ TeV, since it adds nothing to our
conclusions.

In the case of the mass relations among the light quark and lepton masses
(relations (II)), we find that our plots do not depend on $M_t$, therefore we
only display them for a representative value.
Here we do not follow the strategy used in the Standard Model case
({\it i.e.}, we do not vary $m_s$), although we still keep the ratios
$m_d/m_u$ and $m_s/m_d$ fixed.
Instead, we look for that value of $m_s(1~{\rm GeV})$ which gives us the best
agreement for relations (II) both for the low and high $M_{GUT}$ (we cannot
expect exact agreement at one scale, since $m_s/m_d$ does not run).
For instance, we can get the same value of $m_s(1~{\rm GeV})$ in two
different ways, either by demanding that at low $M_{GUT}$ $m_d/3m_e=1.1$ and
$3m_s/m_\mu=1$ or at high $M_{GUT}$ that $m_d/3m_e=1$ and
$3m_s/m_\mu=.9$. In both cases, the masses of the lightest quarks at 1 GeV
are:  $m_u = 3.80\ {\rm MeV}$, $m_d = 6.67\ {\rm MeV}$, and
$m_s = 141~\ {\rm MeV}$.
These results are summarized in Fig.~\ten.

\bigskip
\centerline{\hbox{
                  \psfig{figure=md3me_susy.ps,height=6.5cm}
                 }}
\smallskip
\centerline{\hbox{
                  \psfig{figure=3msmmu_susy.ps,height=6.5cm}
                 }}
\smallskip
\centerline{\singlespace
      \vtop{\parindent=0pt\hsize=5.5truein {\ninerm FIG.~6
Plots of $\scriptstyle m_d/3m_e$ and
$\scriptstyle 3m_s/m_\mu$ as a function of
scale in the SUSY case with $\scriptstyle
M_{SUSY}=1$ TeV, for the low and high unification scales (no appreciable
dependence on $\scriptstyle M_t$ was found).}}}
\bigskip

In the above, our philosophy has been to take
the known low energy data, and using the renormalization group, derive its
implications at high energy.  As we originally did for relation (I),
\refmark{\ARASON}
we could impose both of relations (II) at one unification scale (low or
high $M_{GUT}$).  This would fix $m_s/m_d$ at this scale, and since $m_s/m_d$
and $m_\mu/m_e$ do not run, that would yield $m_s/m_d=23.6$, a value that is
only $12\%$ larger than the value in Refs.~[\gl,\ufgs].  Furthermore, the
results of our runs yield the masses at $1$ GeV of the down and strange
quarks to be $m_d=5.86$ MeV, $m_s=138$ MeV, in the low $M_{GUT}$ case,
and $m_d=6.49$ MeV, $m_s=153$ MeV, in the high $M_{GUT}$ case.  We note that
this approach has also been taken by the authors of Ref.~[\DHR].

Before discussing relation (IV), let us note that, with Supersymmetry,
relation (III) is again well satisfied at all scales.
We now turn to relation (IV). As we did in the Standard Model case we display
our results both for the central value of $V_{cb}$ (.043) and for its upper
value (.050). Then we look for values of $M_t$ which give us agreement at
the unification scale (low or high $M_{GUT}$).  Using the
central value for $V_{cb}$, we find no agreement at the
unification scale. However, this relation is satisfied at Planck scale, if we
use both a high $M_{GUT}$ and $M_t=198$ GeV (the highest possible value
consistent with relation (I)), as displayed in Fig.~\twelve.
\bigskip
\centerline{\hbox{
                  \psfig{figure=vcb44mcmt.ps,height=6.5cm}
                 }}
\smallskip
\centerline{\singlespace
      \vtop{\parindent=0pt\hsize=5.5truein {\ninerm FIG.~7
Plot of $\scriptstyle V_{cb}$ and $\scriptstyle \sqrt{m_c/m_t}$
as a function of scale in the SUSY case
with $\scriptstyle M_{SUSY}=1$ TeV for $\scriptstyle M_t=198$ GeV
and for $\scriptstyle V_{cb}(M_Z)=.043$.}}}
\bigskip
\noindent
We have also
made several runs with a higher value $V_{cb}$.
There, the relation can actually be satisfied  provided that we use  the high
$M_{GUT}$ scale and $M_t=198$ GeV as shown in Fig.~\thirteen
\smallskip
\centerline{\hbox{
                  \psfig{figure=mcmtvcb05.ps,height=6.5cm}
                 }}
\smallskip
\centerline{\singlespace
      \vtop{\parindent=0pt\hsize=5.5truein {\ninerm FIG.~8
Same as Fig.~7 with $\scriptstyle V_{cb}(M_Z)=.050$ and $\scriptstyle M_t=198$
GeV.}}}
\smallskip
\noindent
In the low $M_{GUT}$ case, the two curves
meet closer to the Planck scale.  In fact, theory does not dictate to us the
exact scale at which the $SO(10)$-inspired relation is valid; it could be much
higher than the scale of unification of the Standard Model's gauge
couplings.  To account for this, we now plot, in Fig.~\fourteen,
$V_{cb}$ as a function of $M_t$, assuming that
relation (IV) is valid at $M_{GUT}$, $10M_{GUT}$, and $100M_{GUT}$, and using
the higher value of $g_3(M_Z)$.
\bigskip
\centerline{\hbox{
                  \psfig{figure=last.ps,height=6.5cm}
                 }}
\smallskip
\centerline{\singlespace
      \vtop{\parindent=0pt\hsize=5.5truein {\ninerm FIG.~9
Plot of $\scriptstyle V_{cb}(M_Z)$ as a function of $\scriptstyle M_t$
assuming relation (IV) holds at various scales $\scriptstyle \gsim M_{GUT}$.}}}
\smallskip
\noindent
Given an initial value of $V_{cb}$ at $M_Z$, Fig.~\fourteen\ can be used to
determine the needed $M_t$ (and hence $\beta$) to satisfy relation (IV)
at $M_{GUT}$, $10M_{GUT}$, or $100M_{GUT}$.
We can see from this figure that
as long as $V_{cb}$ is larger than its central value, then relation (IV) can
be satisfied above the $SU(5)$ GUT scale and still allow for a lower value of
$M_t$.

\section{Other Possible Mass Relations}

The relation in Sec.~4.5 involving the
determinants of the charge $-1/3$ and charge $-1$ fermion mass matrices
holds in the minimal supersymmetric model at $10^{14}$ GeV in the high
$M_{GUT}$ case and at $10^{18}$ GeV in the low $M_{GUT}$ case.  In both cases,
this relation holds within $\sim 10\%$ at $10^{16}$ GeV.  These results are
true for the light quark masses chosen in Sec.~5.1 and
for all $160 \leq M_t \leq 198$ GeV.

{\it A priori} one might naively assume that $m_u m_t = m_c^2$ could be
easily satisfied because of the uncertainty in $m_t$.  However, two facts make
the relation viable in the supersymmetric case. First, the value predicted for
the top mass is within the range allowed
by experiment and the $\rho$-parameter bound.  Second, and most remarkable, is
the fact that this top mass value is compatible with relations (I)-(III) given
the choice of light quark
masses in Sec.~5.1. In
Fig.~\fifteen, we display the running of the ratio $m_u m_t / m_c^2$ for the
low and high $M_{GUT}$ cases.
\medskip
\centerline{\hbox{
                  \psfig{figure=mumtmc2.ps,height=6.5cm}
                 }}
\smallskip
\centerline{\singlespace
      \vtop{\parindent=0pt\hsize=5.5truein {\ninerm Fig.~10
Plot of $\scriptstyle m_u m_t / m_c^2$ as a function of scale for the highest
value of $\scriptstyle \alpha_3(M_Z)$ (high curve) and the lowest value of
$\scriptstyle \alpha_3(M_Z)$ (low curve) and for $\scriptstyle M_t=160$ GeV.}}}
\smallskip
\noindent
We show the curves representing the lower $M_t$ value of $160$ GeV for which
the relation is best satisfied at $M_{GUT}$. This relation is incompatible with
relation (IV) however, since the latter favors a higher top mass.
We note that $m_c$ affects these two
relations in an ``inverse'' manner. A lower experimental value for the charm
quark mass favors relation (IV) whereas a higher experimental value favors the
geometric mean relation.

One may also consider the geometric mean relation in the down sector.
We find however that this relation fails to hold in the supersymmetric case.
Other relations among the Yukawa couplings have been considered in the
literature. Theoretical bias or numerology can lead to still other relations
valid at some unifying scale.  In all cases, a thorough renormalization group
analysis will be required in investigations of a possible deeper structure.
\endpage

\chapter{Conclusion}

The aim of this paper has been to explore physics beyond the Standard Model
by studying mass and mixing angle patterns suggested by Grand Unified Theories,
first within the Standard Model context and second in its Minimal
Supersymmetric Extension.  We have reduced some of the parameter space
by constraining the ratios $m_d/m_u$
and $m_s/m_d$, which are better known than the masses themselves.

In the Standard Model case, there are many unsatisfactory features, not the
least of which is the failure of the gauge couplings to unify within
experimental error, forming a GUT triangle.  The simplest of the
$SU(5)$ relations, $m_b=m_\tau$, can only be satisfied some ten orders of
magnitude from the scale of the GUT triangle.  The other
relations $m_d=3m_e$, $3m_s=m_\mu$, $\tan\theta_c=\sqrt{m_d/m_s}$, and
$V_{cb}={\sqrt {m_c/m_t}}$,
could be satisfied at $10^{16}$ GeV.  The geometric mean relations we
considered can also be simultaneously satisfied, but this requires a
top quark mass greater than $200$ GeV.

In the SUSY case, for which the GUT triangle collapses, we achieved a striking
agreement for the four main relations considered.  But, for this to be true,
several things must occur:  first $V_{cb}$ must be larger than its presently
measured value; second the top quark mass must be around $190$ GeV (if it is a
bit lighter, then agreement dictates that $V_{cb}$ should be larger
still); third the Higgs mass should hover around $120$ GeV. These conclusions
are qualitatively correct if one demands maximum agreement.  An analysis
which recently appeared in the literature has reached similar
conclusions.\refmark{\DHR}  However, it is difficult to arrive at more definite
numbers without an exhaustive analysis of the parameter space.

\endpage
\refout

\bye